# Absorbed power in ultracold polarized Fermi mixtures at normal-superfluid separation phase: Mass-imbalanced effect

N. Ebrahimian[1]

[1]Department of Physics, Faculty of Basic Sciences, Shahed University, Tehran, 3319118651, Iran

Email: n.ebrahimian@shahed.ac.ir

## Abstract

Considering ultracold spin-imbalanced Fermi-Fermi mixtures with different spin up and down masses, the absorbed power, subject to an external perturbation with low frequency, has been calculated. The system is composed of spin-up quasiparticles and spin-down quasiholes. The average chemical potential and energy gap have also been numerically calculated via applying different fixed interaction strengths and masses, and then by solving coupled differential equations characterized by Hartree-Fock potential for the spin species, as well as that of the phase separation (PS), leading to imbalance chemical potential. The dependence of the imbalance and average chemical potential in PS regime, to the polarization of the normal component, mass ratios, and interaction strengths are analyzed. Examining density of states (DOS), and by applying the Fermi golden rule at finite temperatures, the absorbed power has accordingly been calculated as a function of temperature, interaction strength, and mass ratio. Finally, the behavior of absorbed power versus frequency has been investigated.

## 1. Introduction

Ultracold Fermi gas is a powerful model for investigating properties of interacting many-body systems [1-3]. Bardeen-Cooper-Schrieffer (BCS) superfluidity to Bose-Einstein condensation (BEC) crossover is a theoretical topic advanced originally by Eagles and Leggett, proposing that the BCS wavefunction is more general than being applicable only to weakly interacting systems [4-5]. Experimental realization of BCS-BEC crossover with a Fermi gas of atoms [6] has showed that a Feshbach resonance[7-17] could change the interaction strength. A magnetic-field Feshbach resonance could tune the s-wave scattering length by varying the strength of magnetic field, leading to a number of conceptual advances. The importance of ultracold Fermi gas lies in its ability to regulate interaction between particles using Feshbach resonance [13]. In such systems, there also is a pseudogap phenomenon, which plays a role in affecting physical quantities of the system. Ultracold spin-polarized(imbalanced) Fermi gas has so far garnered a lot of attention from both experimental and theoretical sides [18-26]. In a spin-balanced Fermi gas, BCS-BEC crossover can be achieved at low temperatures when scattering between atoms is tuned via the resonance; therefore, the system can evolve smoothly from BCS to BEC regime [27]. It should be noted that when the low-energy s-wave scattering length, $a$, and the interaction strength, $1/k_F a$ are negative, the two atomic species accordingly interact weakly with each other, and the superfluid phase is also formed by weakly bound Cooper pairs; this situation is referred to the BCS regime. On the contrary, when $a > 0$, the system falls into the Bose-Einstein condensation state. However, a spin- polarized Fermi gas might associate with phase separation (PS) between normal and superfluid phases at very low temperatures [28-35]. The phase separation takes place when the chemical potential difference between two spin species or imbalance chemical potential, $h_s$ (assuming $h_s > 0$) approaches to a critical value, $h_c$. Such a phase-separation scenario had long been proposed by Clogston and Chandrasekhar[36-37]. At zero temperature and in the BCS limit, they found that $h_c = \Delta/\sqrt{2}$, with $\Delta$ is the energy gap. The conditions for normal–superfluid phase separation are established as follows. Phase separation occurs only under specific thermodynamic conditions. First, the chemical potentials of each species must be equal in the superfluid and normal phases. If, for the spin-up species, this equality did not hold, quasiparticles would move from the phase with the higher value to the phase with the lower value, lowering the grand-canonical potential and violating the assumed thermodynamic equilibrium. Second, the grand-canonical thermodynamic potentials of the superfluid and normal phases must be equal. Third, the pressures of the two phases must be equal at the interface. When the superfluid and normal phases are homogeneous, the pressure is directly determined by the grand-canonical potential, so these conditions are mutually consistent. The phase diagram of the spin-polarized Fermi gas was also theoretically studied by different methods such as pairing fluctuation theory [38-40]. Some works have investigated the normal-superfluid separation regime on the BCS side of BCS-BEC crossover, including the study of interface thermal conductivity [32,41-42]. The overwhelming majority of experiments on polarized Fermi gases have been done at equal masses. However, theoretical investigations have predicted fermionic systems with mass imbalance to favor exotic interaction regimes [43-45]. Recently, the realization of a Fermi-Fermi mixture of ultracold atoms, such as $^{161}$Dy and $^{40}$K, have also been investigated in Refs. [43-45]. Also, comprehensive discussions and detailed investigations on ultracold Fermi gases, including their phase diagrams and thermodynamic properties across different regimes such as normal–superfluid phase separation, as well as critical and transition temperatures, can be found in Refs. [2,17,30-31,40,46-72].

The present work focuses on a polarized Fermi-Fermi mixture, consisting of two spin species, though with unequal masses. The allowed numerical values of the imbalance and average chemical potentials, energy gap, and the mass ratio, $m_r$, have accordingly been calculated within the BCS regime with various interaction strengths. Basically, the mass asymmetry is a feature of Fermi-Fermi mixtures, such as $^{40}$K-$^{6}$Li mixture with mass ratio 6.7. By considering the occurrence of normal-superfluid separation in the BCS regime in a Fermi-Fermi mixture, the present work accordingly delves into the absorbed power as a function of temperature, imbalance chemical potential, interaction strength and frequency. Different mixtures with different masses have also been taken into account. It should be noted that in such mixtures, the Clogston relation between imbalance chemical potential and energy gap cannot be applicable; rather, imbalance chemical potential has been calculated by equating the related grand canonical thermodynamic potentials at superfluid and normal components, as is explained later on.

In this paper, we first determine the relevant parameters required to calculate the response of the system to the ultrasound wave. Therefore, in PS regime of the BCS side, the imbalance and average chemical potentials versus the polarization of the normal component, interaction strength, and mass ratios are analyzed.

The dynamic response of a BCS superconductor to a time-dependent external perturbation was investigated[73-74]. Response to sound waves, ultrasound waves, microwaves (or electromagnetic waves) can be examples of the applied time-dependent external perturbation to the system. An time-dependent external perturbation induces transitions between different excited states which leads to attenuation or delay of the perturbation. By evaluating the transition probabilities associated with the perturbation, one can deduce the relative change in the decay rates bouroght on by superfluidity[74]. It has been known that in the presence of normal-superfluid separation in BCS regime for Fermi-Fermi mixtures, there is no theoretical information about absorbed power; therefore, one of the main purposes of our work is to unveil the behavior of different Fermi-Fermi mixtures in normal-superfluid separation regime as the mass ratio of up and down spins changes. Regarding the mass ratio, one of our results is close to the $^{161}$Dy-$^{40}$K mixture, since its mass ratio is about 4.08. Moreover, the results of this paper may be extended to solid-state systems, for example, to certain transition-metal dichalcogenides.

The present paper is organized as follows. In section 2, first, the Hamiltonian of the system is introduced and Bogoliubov coefficients is given. Then, the investigation of the average and imbalance potentials in terms of the polarization in the normal component, mass ratios, and interaction strengths in the PS regime is presented. In section 3, the Perturbed Hamiltonian due to ultrasonic wave is given and density of states (DOS) are introduced and calculated. Then, after the formulation of the absorbed power, the results are given and discussed in terms of temperature and relevant parameters. We conclude our discussion in section "Conclusions".

## 2. Thermodynamic Parameters
## 2-I. Method of Calculations

The effective Hamiltonian of a spin-polarized Fermi gas consisting of two fermionic species of masses $m_\uparrow$ and $m_\downarrow$ and chemical potentials $\mu_\uparrow$ and $\mu_\downarrow$, is given by [32,41] (we work in Planck units; $k_B = \hbar = 1$)

$$\hat{H} = \int d^3x \sum_i [\hat{\psi}^\dagger(\vec{r}\,i)H(\vec{r}\,i)\hat{\psi}(\vec{r}\,i) + U(\vec{r}\,i)\hat{\psi}^\dagger(\vec{r}\,i)\hat{\psi}(\vec{r}\,i)$$
$$+ \Delta(\vec{r})\hat{\psi}^\dagger(\vec{r}\uparrow)\hat{\psi}^\dagger(\vec{r}\downarrow) + \Delta^*(\vec{r})\hat{\psi}(\vec{r}\downarrow)\hat{\psi}(\vec{r}\uparrow)] \qquad (1)$$

where $\psi$ and $\psi^\dagger$ are annihilation and creation field operators, respectively. $H(\vec{r}\,i)$ is

$$H(\vec{r}\,i) = -(1/2m_i)(\vec{\nabla} - (ie\vec{A}/c))^2 - \mu_i \qquad (2)$$

where $i = \uparrow, \downarrow$, $\vec{A}$ is vector potential ($\vec{A}$ equal to zero), $\mu_i$ and $m_i$ are chemical potential, and mass for spin $i$, respectively, and $\Delta(\vec{r})$ is energy gap and it is assumed as a real function, i.e. $\Delta(\vec{r}) = \Delta^*(\vec{r}) = \Delta$. $U(\vec{r}\,i)$ is the Hartree-Fock (HF) potential for spin $i$ and is given by $U(\vec{r}\downarrow) = V\langle \psi^\dagger(\vec{r}\uparrow)\psi(\vec{r}\uparrow)\rangle$ and $U(\vec{r}\uparrow) = V\langle \psi^\dagger(\vec{r}\downarrow)\psi(\vec{r}\downarrow)\rangle$ where $\langle...\rangle$ is quantum expected value. The interaction between up and down spins is also assumed to be a contact interaction characterized by the coupling constant $V = -4\pi a/m_+$ with $m_+ = 2m_\uparrow m_\downarrow/(m_\uparrow + m_\downarrow)$. Since, in the BCS limit, the two species of particles with spin up and spin down in superfluid phase are of equal densities, which means that there is no magnetized superfluid phase that is thermodynamically stable. then $U(\vec{r}\downarrow) = U(\vec{r}\uparrow) \equiv U_s$ where $U_s$ is the HF potential in superfluid component of the system, however, the normal phase can be polarized, and there is the phase-separated region in which the superfluid and normal states are spatially separated and in which metastable spin-polarized superfluid solutions can appear. Therefore, in the normal phase, then one has

$U(\vec{r}\downarrow) \neq U(\vec{r}\uparrow)$. It should be noted that although the occurrence of superfluid–normal phase separation does not strictly require the inclusion of HF potential, accounting for them leads to more realistic results. This is because these potentials appear in the thermodynamic potential matching condition, which is one of the key criteria for phase separation. Therefore, the HF potential plays an essential role in ensuring proper pressure matching between phases, determining the thermodynamic potentials of the superfluid and normal components, and maintaining the stability of the two-phase system. The energy gap, $\Delta(\vec{r})$, would also be independent of $\vec{r}$ assuming a homogeneous system of two fermion species at fixed chemical potentials $\mu_\uparrow$ and $\mu_\downarrow$, without any external potential. By considering field creation (annihilation) operator in terms of fermionic creation (annihilation) operator, $a_{k,\sigma}^\dagger (a_{k,\sigma})$ with momentum vector $\vec{k}$ and spin $\sigma$, as $\psi_\sigma^\dagger(\vec{r}) = \sum_{\vec{k}} e^{i\vec{k}\cdot\vec{r}} a_{k,\sigma}^\dagger$ ($\psi_\sigma(\vec{r}) = \sum_{\vec{k}} e^{i\vec{k}\cdot\vec{r}} a_{k,\sigma}$), one can write

$$H = \sum_{\vec{k},\sigma} \left( \frac{k^2}{2m} - \mu_\sigma + U_\sigma \right) a_{\vec{k},\sigma}^\dagger a_{\vec{k},\sigma} + \sum_{\vec{k}} \Delta \left( a_{\vec{k},\uparrow}^\dagger a_{-\vec{k},\downarrow}^\dagger + a_{-\vec{k},\downarrow} a_{\vec{k},\uparrow} \right) \quad (3)$$

where $U_\sigma$ is the Fourier transform of $U(r\sigma)$. Also, $H$ given by Eq. (1) can be written in terms of Bogoliubov quasiparticle (or fermionic quasiparticle) creation and annihilation operators, $\gamma^\dagger$ and $\gamma$, when the following general Bogoliubov transformation and its Hermitian adjoint were used[73-74]

$$\psi_\sigma(\vec{r}) = \sum_{\vec{k}} u_{\vec{k},\sigma}(\vec{r}) \gamma_{\vec{k},\sigma} - v_{\vec{k},\sigma}(\vec{r}) \gamma_{\vec{k},-\sigma}^\dagger \quad (4)$$

where $u_{\vec{k},\sigma}(\vec{r})$ and $v_{\vec{k},\sigma}(\vec{r})$ quasiparticle wave functions that are satisfied in Bogoliubov equations. The result is $H = \sum_{\vec{k},\sigma} E_{\vec{k}_{\alpha,\beta}} \gamma_{\vec{k},\sigma}^\dagger \gamma_{\vec{k},\sigma}$ where $E_{\vec{k}_{\alpha,\beta}}$ is the excitation of energy. The $\alpha$ ($\beta$) branch is composed of spin-up (-down) quasiparticles $\uparrow$ and spin-down (-up) quasiholes $\downarrow$. In this paper $\alpha$ branch is considered. The average and imbalance chemical potential is defined as $\mu_s = ((\mu_\uparrow + \mu_\downarrow)/2) - U_s$ and $h_s = (\mu_\uparrow - \mu_\downarrow)/2$, respectively. The excitation energy, $E_{\vec{k}_{\alpha,\beta}}$, of the system in terms of $\mu_s$ and $h_s$ is then [32]

$$E_{\alpha,\beta} = \mp h_s \pm \frac{k^2}{2m_-} + \sqrt{\varepsilon_{p_{\alpha,\beta}}^2 + \Delta^2} \qquad (5)$$

where $\varepsilon_{\vec{k}_\alpha} = \varepsilon_{\vec{k}_\beta} \equiv (k^2/2m_+ - \mu_s)$ with wave vector $\vec{k}$ and $m_- = 2m_\uparrow m_\downarrow/(m_\downarrow - m_\uparrow)$.

To calculate the absorbed power of the polarized Fermi gas, the values of $h_s, \mu_s$ and $\Delta$ are first required, being numerically calculated as follows. The temperature dependence of the energy gap is $\Delta - \sqrt{2\pi T \Delta}\left(1 - (T/8\Delta)\right)e^{-\Delta/T}$ [75]. At low temperatures, the change of the energy gap with respect to temeperature is small and can be ignored and the energy gap can be considered constant (this expression refers to equal masses; the case of unequal masses will be discussed later, where the temperature dependence can likewise be neglected).

Now one needs numerically calculate the relevant parameters, i.e. average chemical potential, energy gap, and finally imbalance chemical potential. For these purposes, the integral equations of the energy gap, HF potential of superfluid component, $U_s$, Fermi wave number, which is given in terms of number density, $n_s$, and the superfluid and normal grand canonical potentials per unit volume. Gap equation is given by $\Delta = -V \sum u_k v_k$. Minimizing the energy of the unperturbed system, the coefficeints, $u_{\vec{k}}$ and $v_{\vec{k}}$, can then be calculated and is given as follows

$$v_{\vec{k}} = \frac{1}{\sqrt{2}} \sqrt{1 - \frac{k^2/2m_+ - \mu_s}{E_{\vec{k}_\alpha} + h_s - k^2/2m_-}} \qquad u_{\vec{k}} = \frac{1}{\sqrt{2}} \sqrt{1 + \frac{k^2/2m_+ - \mu_s}{E_{\vec{k}_\alpha} + h_s - k^2/2m_-}} \qquad (6)$$

Using the coefficients, gap equation becomes as follows

$$1 = \frac{-V}{2} \int \frac{d^3 k}{(2\pi)^3} \left( \frac{1}{\sqrt{\left(\frac{k^2}{2m_+} - \mu_s\right)^2 + \Delta^2}} - \frac{1}{\frac{k^2}{2m_+}} \right) \qquad (7)$$

It should be noted that in Eq. (7), the contribution of normal state was substructed. Then, by eleminating the ultraviolet divergencies[76], changing the variables $\left(k^2/2m_+\mu_s \to z,\ \Delta/\mu_s \to X\right)$, defining $\varsigma = -\left[1+X^2\right]^{-1/2}$ and using the following relation[76]

$$\int_0^\infty \frac{z^\delta dz}{\sqrt{(z-1)^2 + X^2}} = \frac{-\pi}{\sin \pi\delta} \varsigma^{-\delta} P_\delta(\varsigma) \tag{8}$$

where $P_\delta$ is the polynomial Legendre of degree $\delta$, one get

$$\frac{1}{m_+ a^2} = -2\mu_s \frac{P_{1/2}^2(\varsigma)}{\varsigma} \tag{9}$$

The combination of Eq. (9) and the definition $\varsigma = -\left[1+(\Delta/\mu_s)^2\right]^{-1/2}$ obtains

$$1 = \varsigma^2 + 4\left(\Delta\, m_+ a^2\right)^2 \left[P_{1/2}(\varsigma)\right]^4 \tag{10}$$

The another needed equation is Fermi wave number, i.e.

$$k_F^3 = 3\pi^2 n_s = 3\pi^2 \int_0^\infty \frac{d^3k}{(2\pi)^3}\left(1 - \frac{k^2/2m_+ - \mu_s}{\sqrt{\left(k^2/2m_+ - \mu_s\right)^2 + \Delta^2}}\right)$$

$$= -\frac{3}{2}\sqrt{2} m_+^{3/2} \mu_s^{3/2} \left(\pi\left(\varsigma^{-2}\right)^{3/4} P_{3/2}(\varsigma) + \pi\left(\varsigma^{-2}\right)^{1/4} P_{1/2}(\varsigma)\right) \tag{11}$$

The combination of Eqs. (9) and (11) and the use of the definition, $\varsigma$ one gets

$$k_F a = (3\pi/4)^{1/3}\left[\left(P_{3/2}(\varsigma)/P_{1/2}(\varsigma) - \varsigma\right)/P_{1/2}^2(\varsigma)\right]^{1/3} \tag{12}$$

Now, from Eq. (12), one can determine the allowed values of $\varsigma$ via fixing $1/k_F a$ in the BCS regime. Furthermore, substituting $a$ from Eq. (12) and $k_F$ from Eq. (11) in Eq. (10), one obtains

$$1 = \zeta^2 + 4\left[\frac{\Delta}{\mu_s} \frac{((3\pi/4))\left[\left(P_{3/2}(\zeta)/P_{1/2}(\zeta)-\zeta\right)/P_{1/2}^2(\zeta)\right]^{2/3}}{-\left(\frac{3}{2}\right)^{2/3} 2^{1/3}\left(\pi\left(\zeta^{-2}\right)^{3/4} p_{3/2}(\zeta)+\pi\left(\zeta^{-2}\right)^{1/4} p_{1/2}(\zeta)\right)^{2/3}}\right]^2 [P_{1/2}(\zeta)]^4 \quad (13)$$

By having $\zeta$, Eq. (13) and the definition $\zeta = -\left[1+(\Delta/\mu_s)^2\right]^{-1/2}$ lead to obtain $\Delta$ and $\mu_s$. One can obtain HF potential of superfluid component, $U_s$, from the following equation

$$U_s = -\frac{V}{2}\int_0^\infty \frac{d^3k}{(2\pi)^3}\left(1 - \frac{k^2/2m_+ - \mu_s}{\sqrt{(k^2/2m_+ - \mu_s)^2 + \Delta^2}}\right) = \mu_s\left(1 - \zeta^{-1}P_{3/2}(\zeta)/P_{1/2}(\zeta)\right) \quad (14)$$

In the last line of Eq. 14, the relation $V = \frac{4\pi a}{m_+}$ and Eq. (9) were used. The imbalance chemical potential, $h_s$, is determind as follows. By considering $\mu_\uparrow = \mu_s + U_s + h_s$ and $\mu_\downarrow = \mu_s + U_s - h_s$, one can write the following HF potentials in the normal phase, $U_\uparrow$ and $U_\downarrow$,

$$U_\uparrow = \frac{a}{m_+}\left(-\frac{4\sqrt{2}}{3\pi}\right)\left(m_r m_\uparrow\left(\mu_s + U_s - h_s - U_\downarrow\right)\right)^{3/2}, U_\downarrow = \frac{a}{m_+}\left(-\frac{4\sqrt{2}}{3\pi}\right)\left(m_\uparrow\left(\mu_s + U_s + h_s - U_\uparrow\right)\right)^{3/2} \quad (15)$$

where $U_\uparrow$ and $U_\downarrow$ are coupled. By imposing equality of the grand-canonical thermodynamic potentials of the superfluid and normal phases, and by requiring that the chemical potentials of each species be the same in both phases, which together provide the conditions for the occurrence of phase separation, one has

$$\frac{3\sqrt{2}\pi\zeta^2}{64|P_{1/2}(\zeta)|^5}\left(\left(\zeta^{-2}-5\right)P_{1/2}(\zeta)+4\zeta^{-1}P_{3/2}(\zeta)\right) =$$
$$m_+ a^5 \left(m_\uparrow\right)^{3/2}\left(\left(\mu_s + U_s + h_s - U_\uparrow\right)^{5/2} + \left(m_r\right)^{3/2}\left(\mu_s + U_s - h_s - U_\downarrow\right)^{5/2}\right) \quad (16)$$

By simultaneously Eqs. (15) and (16), $U_\uparrow$, $U_\downarrow$ and $h_s$ can be obtained. Since $h_s$ is a real and $U_\uparrow$, $U_\downarrow$, and $U_s$, are negative, numerical computations indicate that only certain mass-ratio values are admissible; this point will be addressed elsewhere.

## 2-II. Systematic Analysis of Thermodynamic Parameters

When a two-component ultracold Fermi gas with s-wave interactions has nonzero spin polarization, one spin species has a larger population than the other, and their Fermi surfaces no longer match. This mismatch makes pairing more difficult; as a result, fewer pairs form than in the unpolarized case. Once pairing begins, all or part of the minority-spin population become paired. The existence of unpaired species, alongside the paired species, increases the energetic cost of maintaining a uniform paired state. If this cost exceeds the condensation gain, the homogeneous superfluid becomes unstable. Keeping the unpaired species mixed with the pairs further raises the energy, both through their interactions with the pairs and because the system would be lower in energy if those quasiparticles and quasiholes could also pair. The system therefore reduces its energy when the unpaired species leave the paired region, which leads to the phase separation into an unpolarized superfluid and a partially or fully polarized normal phase (here we focus on the partially polarized case) . The phase-separated state remains stable as long as the normal and superfluid components satisfy the standard thermodynamic equalities: equal pressure and equal chemical potentials for each spin species.

Meanwhile, mass imbalance strengthens the inhibition mechanism of pairing formation by increasing the mismatch between the two spin components' Fermi surfaces and by altering the density of states for each component. As a result, the stability threshold of the uniform superfluid shifts (typically becoming less robust when the heavy component is the majority), and the unpaired population, present alongside the pairs, further raises the free energy of a uniform, polarized condensate compared with the equal-masses case.

In the following, some results of the numerical calculations for the relevant parameters are presented. In Fig. 1, the imbalance chemical potential, $h_s$, at which normal–superfluid phase separation occurs, is plotted as a function of the interaction strength for three different mass ratios.

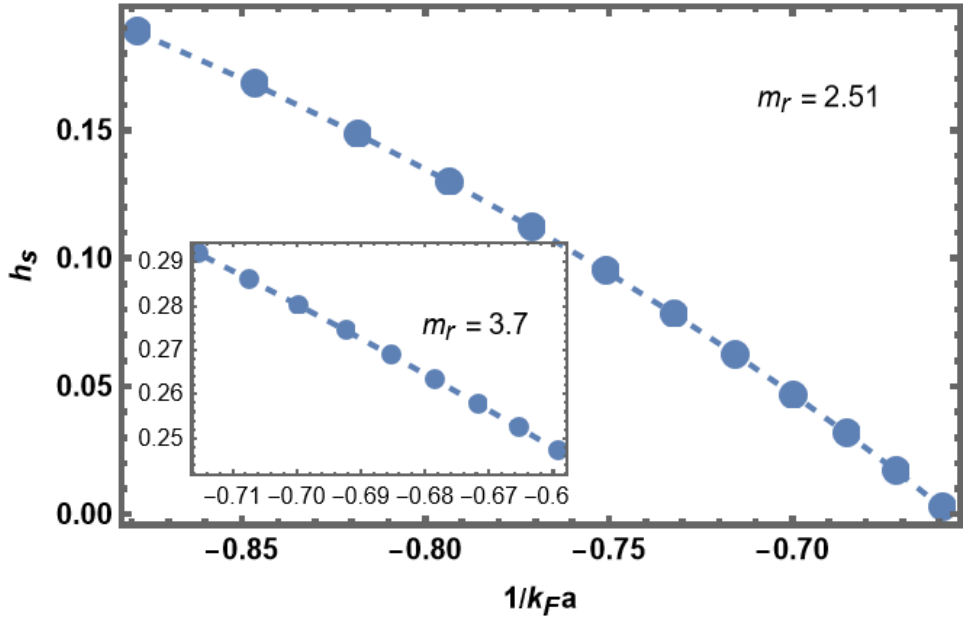

Fig.1: (Color online) the imbalance chemical potential (measured with respect to $T_F$) at which normal–superfluid phase separation occurs as a function of the interaction strength for a fixed mass ratio, $m_r = 2.51$ ( Inset: For $m_r = 3.7$ ).

Also, Fig. 2 shows the imbalance chemical potential, $h_s$, at which normal–superfluid phase separation occurs, plotted as a function of the mass ratio for two fixed interaction strengths, $1/k_F a \simeq -0.659$, and $-0.878$.

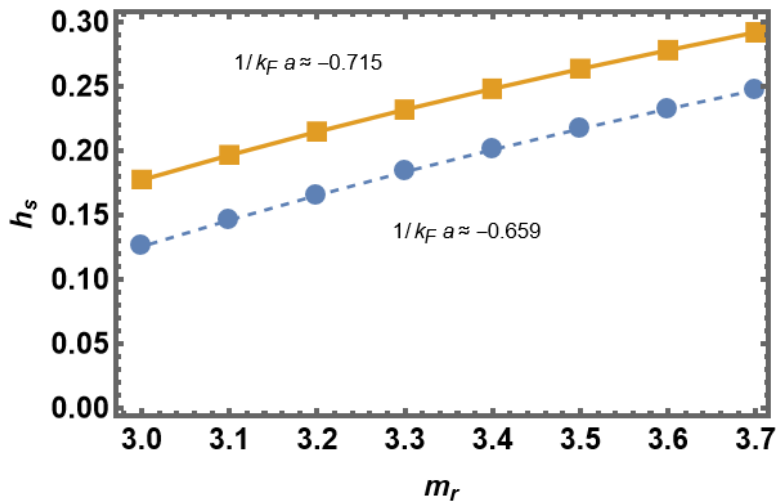

Fig.2: (Color online) the imbalance chemical potential, $h_s$, (measured with respect to $T_F$) at which normal–superfluid phase separation occurs, as a function of the mass ratio for two fixed interaction strengths, $1/k_F a \simeq -0.659$, and $-0.878$.

It is seen that as the mass ratio increases (i.e., the spin-down component becomes heavier), the imbalance chemical potential required for normal–superfluid phase separation increases. Likewise, increasing the magnitude of the interaction strength raises the imbalance chemical potential. Hence, $h_s$ depends on both the interaction strength and the mass ratio. By contrast, Figs. 3 and 4 show that, on the BCS side within the normal–superfluid phase-separation regime, the average chemical potential $\mu_s$ and $\Delta$ is essentially independent of the mass ratio and varies only with the interaction strength. Although the average chemical potential (similar argument holds for the energy gap) is identical for all mass ratios at a given interaction strength, the admissible range of interaction strength depends on the mass ratio because the condition $h_s < \Delta$ must be satisfied. Accordingly, Figs. 3 and 4 are presented for the case of mass ratio, $m_r = 2.51$.

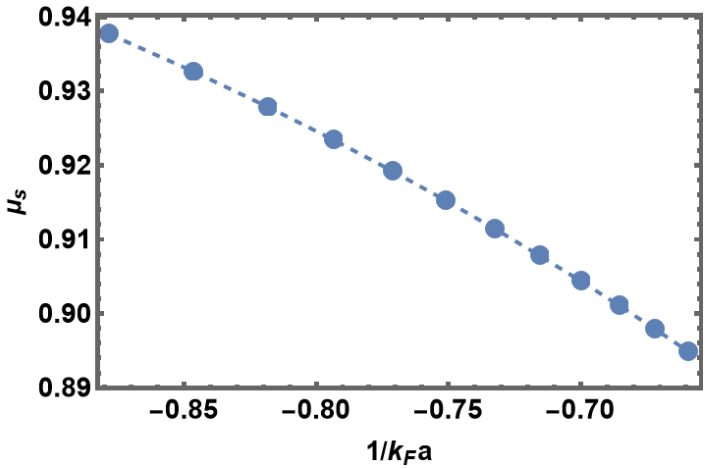

Fig.3: (Color online) The average chemical potential (in $k_B T_F$ units) on the BCS side within the normal–superfluid phase separation regime, as a function of the interaction strength. Although the average chemical potential is identical for all mass ratios at a given interaction strength, the admissible range of interaction strength depends on the mass ratio because the condition $h_s < \Delta$ must be satisfied. Accordingly, the figure is presented for the case of mass ratio, $m_r = 2.51$.

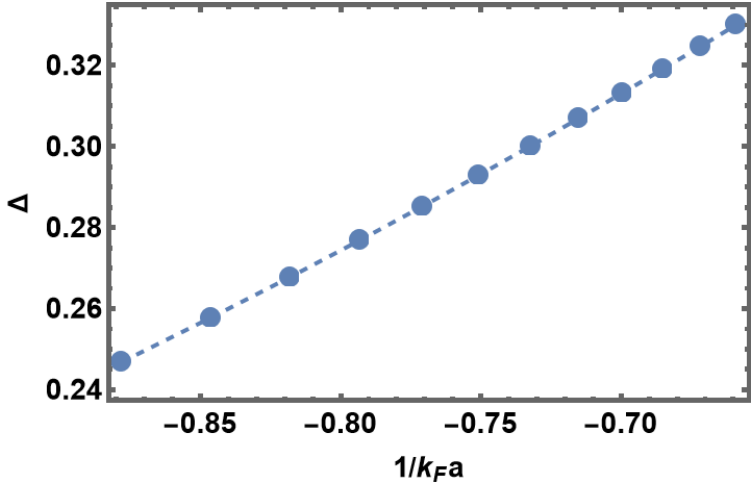

Fig.4: (Color online) The energy gap (in $k_B T_F$ units) on the BCS side within the normal–superfluid phase separation regime, as a function of the interaction strength. Although the energy gap is identical for all mass ratios at a given interaction strength, the admissible range of interaction strength depends on the mass ratio because the condition $h_s < \Delta$ must be satisfied. Accordingly, the figure is presented for the case of mass ratio, $m_r = 2.51$.

Now the relation between the spin polarization in normal component and other relevant parameters is considered. Polarization in the normal component can be calculated by the following relation

$$p_n = \frac{n_\uparrow^n - n_\downarrow^n}{n_\uparrow^n + n_\downarrow^n} \qquad (17)$$

where $n_\delta^n$ denotes the number density of spin-$\delta$ in the normal phase and is given by $n_\delta^n = \int d^3k/(2\pi)^3 = \int N(E_i) dE_i$ (the volume is set to one) where $N(E_i)$ is density of states of the normal phase. Here, the energy spectrum of the normal case for each spin, $E_i$, is

$$E_\uparrow = \left| \frac{k^2}{2m_\uparrow} - \mu_s - h_s - U_s + U_\uparrow \right| \quad , \quad E_\downarrow = \left| \frac{k^2}{2m_\downarrow} - \mu_s + h_s - U_s + U_\downarrow \right| \qquad (18)$$

Then the polarization of the normal case is given by

$$p_n = \frac{(\mu_s + h_s + U_s - U_\uparrow)^{3/2} - (m_r)^{3/2}(\mu_s - h_s + U_s - U_\downarrow)^{3/2}}{(\mu_s + h_s + U_s - U_\uparrow)^{3/2} + (m_r)^{3/2}(\mu_s - h_s + U_s - U_\downarrow)^{3/2}} \qquad (19)$$

As noted earlier, before any phase separation occurs, unpaired quasiparticles or quasi-holes within the superfluid component raise its energy and can even make it unstable. For example, when spin polarization is very high, the mismatch between the two Fermi surfaces can become so large that pairing is no longer possible. Conversely, when spin polarization is very small, the probability of unpaired quasiparticles or quasiholes in the outer layer decreases. When the masses are unequal, pairing between the two species can become more difficult, so fewer pairs form. The conditions for phase separation also change, and phase separation may occur at lower values of polarization than in the equal-mass case.

In what follows, we analyze spin polarization and its relation to the thermodynamic parameters, the interaction strength, and the mass ratio.

In Fig. 5, the polarization of the normal component is shown as a function of the mass ratio. At fixed interaction strength, a larger mass ratio reduces the polarization magnitude; at fixed mass ratio, stronger interactions likewise reduce the polarization. Fig. 6 shows polarization versus interaction strength for two mass ratios, $m_r = 2.51$ and 3.7. In Fig. 7 and its inset, the imbalance and average chemical potentials are plotted as functions of polarization at a fixed mass ratio, $m_r = 3.7$, respectively. At a fixed mass ratio, stronger attraction (i.e., an increase in the magnitude of the interaction strength) forms more pairs and lowers the superfluid free energy. To maintain coexistence, the system then requires a smaller imbalance chemical potential; the Fermi surfaces of the two spin components move closer together, so its polarization decrease.

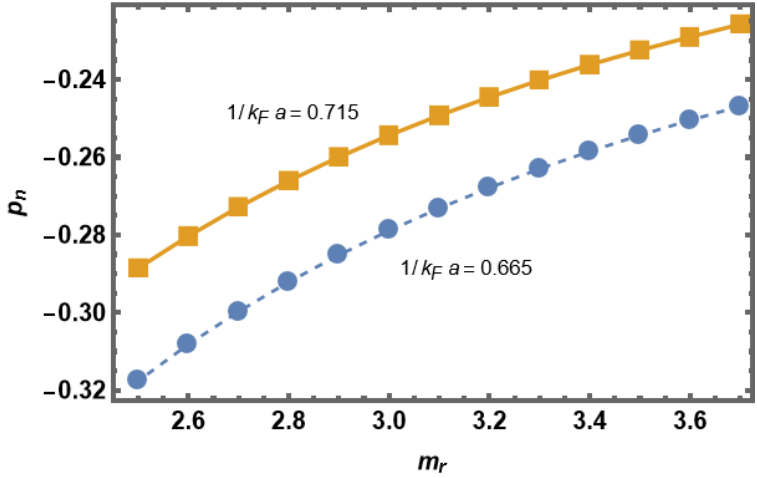

Fig. 5: (Color online) The spin polarization of the normal component on the BCS side within the normal–superfluid phase separation regime, as a function of the mass ratio at two different interaction strengths, $1/k_F a = -0.878$, and -0.672.

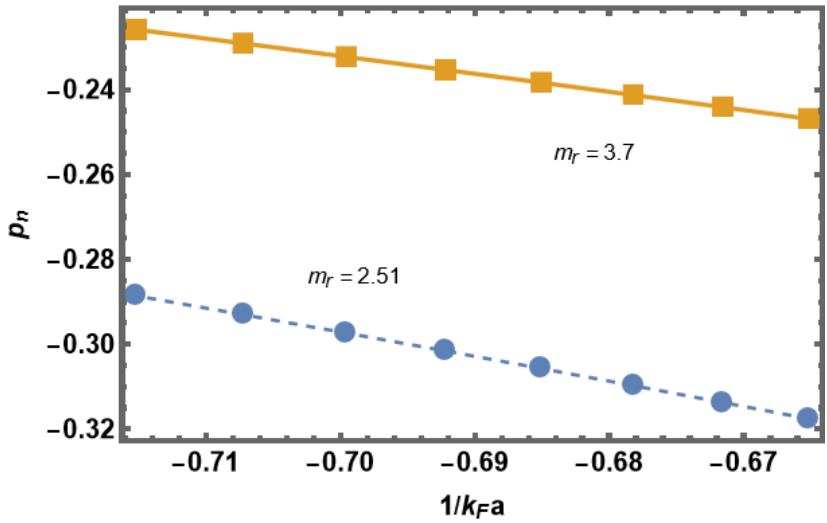

Fig. 6: (Color online) The spin polarization of the normal component on the BCS side within the normal–superfluid phase separation regime, as a function of the interaction strength at two different mass ratios, $m_r = 2.51$ and 3.7.

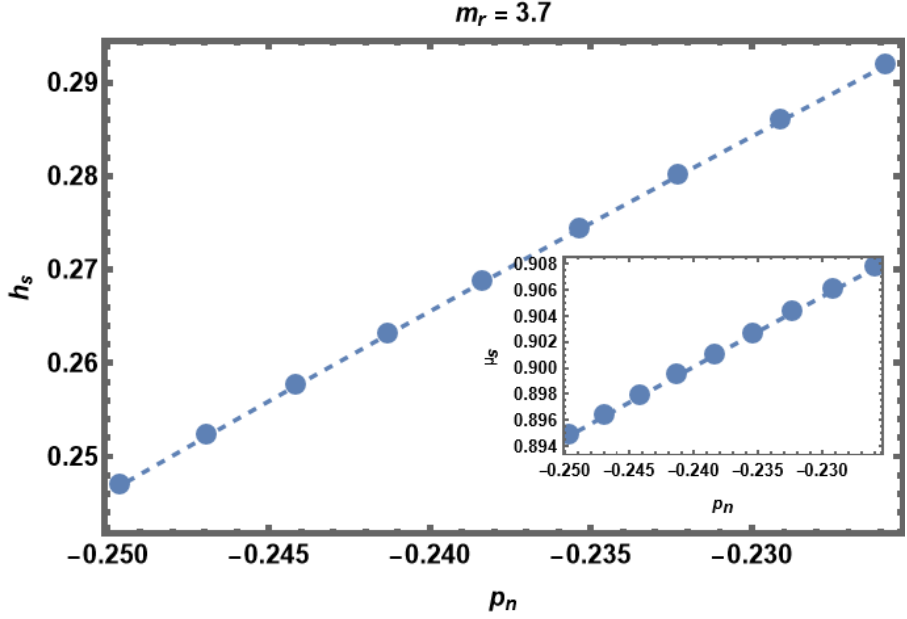

Fig. 7: (Color online) The imbalance chemical potential in terms of the spin polarization of the normal component on the BCS side within the normal–superfluid phase separation regime at a fixed mass ratio, $m_r = 3.7$. Inset: The average chemical potential in terms of the in terms of the spin polarization of the normal component on the BCS side within the normal–superfluid phase separation regime at a fixed mass ratio, $m_r = 3.7$.

Also, the total polarization can be obtained via

$$p = \frac{n_\uparrow^n - n_\downarrow^n}{n_\uparrow^n + n_\downarrow^n + n_s} \qquad (20)$$

where the number density of the pairs in unpolarized superfluid component, $n_s$, is given by Eq. (11). It is worth mentioning that, in Figures 5–7, when $p_n$ is replaced by $p$, the figures exhibit only a shift. This occurs because the superfluid component is unpolarized. Accordingly, only a single figure is presented below: the polarization as a function of interaction strength (Fig. 8).

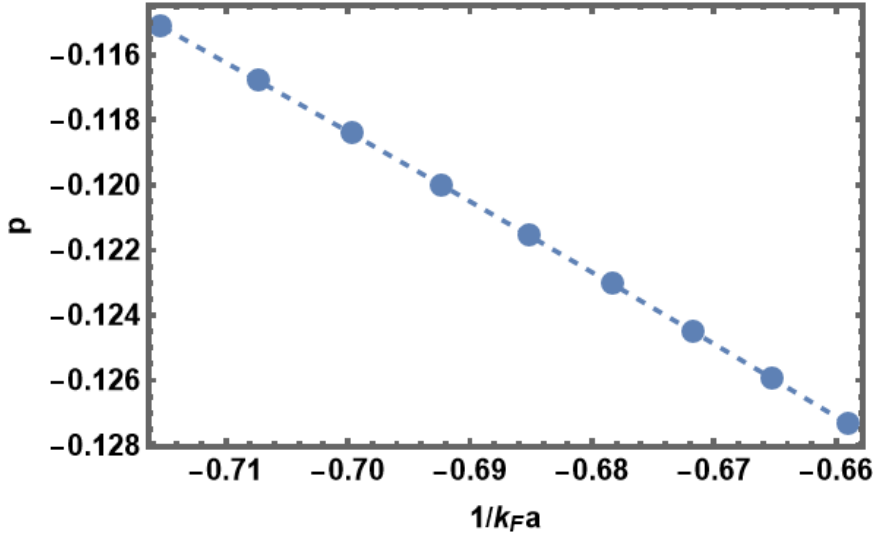

Fig. 8: (Color online) The total polarization, $p$, and the polarization of the normal component, $p_n$, on the BCS side within the normal–superfluid phase separation regime, as a function of the interaction strength at a fixed mass ratio, $m_r = 3.7$.

The equations given above for $\Delta$, $\mu_s$, and $h_s$, and also, the grand-canonical thermodynamic potentials of the superfluid and normal components were written at zero temperature. As long as the conditions $T \ll \Delta$, $T \ll \mu_\uparrow - U_\uparrow$, and $T \ll \mu_\downarrow - U_\downarrow$ holds, they provide a good approximation at finite temperatures with a small error. For example, the finite-temperature gap equation is given below

$$1 = \frac{-V}{2}\int \frac{d^3k}{(2\pi)^3}\left(\frac{1-f(E_\alpha)-f(E_\beta)}{\sqrt{\left(\frac{k^2}{2m_+}-\mu_s\right)^2+\Delta^2}} - \frac{1}{\frac{k^2}{2m_+}}\right) \qquad (20)$$

where $f$ is Fermi–Dirac distribution function. Under the condition, $T \ll \Delta$, its contribution are small, so the equation can be approximated by its zero-temperature form. In the normal component, for example, for the spin-up particle density at finite temperature

$$n_\uparrow^n = \int d^3k/(2\pi)^3 f(E_\uparrow) = \int d^3k/(2\pi)^3 f\left(\frac{k^2}{2m_\uparrow}-\mu_\uparrow+U_\uparrow\right) \qquad (21)$$

when $T \ll \mu_\uparrow - U_\uparrow$, only a negligible number of particles occupy states above the level $\mu_\uparrow - U_\uparrow$. The Fermi-Dirac function in Eq. (21) is then essentially one, and the particle density equals its $T = 0$ value. Throughout the system under study, these conditions hold; therefore we use the zero-temperature equations. To verify this, the maximum temperature considered is $0.02 T_F$, and the pairing gap in Fig. 4 never falls below $0.2 T_F$, hence $T \ll \Delta$. Moreover, using $\mu_\uparrow - U_\uparrow \equiv \mu_s + h_s + U_s - U_\uparrow$ where $U_\uparrow$ is the HF potential in the normal component, Fig. 9 shows that $T \ll \mu_\uparrow - U_\uparrow$ (similar argument holds for $T \ll \mu_\downarrow - U_\downarrow$; see also Fig. 10).

To determine the critical temperature that is, the transition from the phase-separated superfluid–normal state to a fully normal state, one must solve the finite-temperature equations self-consistently, a nontrivial task that we defer to future work. Here, however, a qualitative estimate of this temperature is provided to verify that the chosen parameters remain within the phase-separated regime. For each mass ratio and interaction strength, the critical temperature can be given for every value of the polarization[51,53,77-78]. Lower spin polarization reduces the critical temperature. At the smallest polarization we examine, the critical temperarure is about $0.025 T_F$ [51,53,77-78]. In contrast, a smaller mass ratio (Fig. 5) or a stonger interaction strength (Figs. 6 and 8) both increase the critical temperature because of the increase of the polarization. Therefore, by working at temperatures below $0.02 T_F$, we remain safely below the critical temperature for all parameter sets, and the normal–superfluid phase-separated regime is preserved.

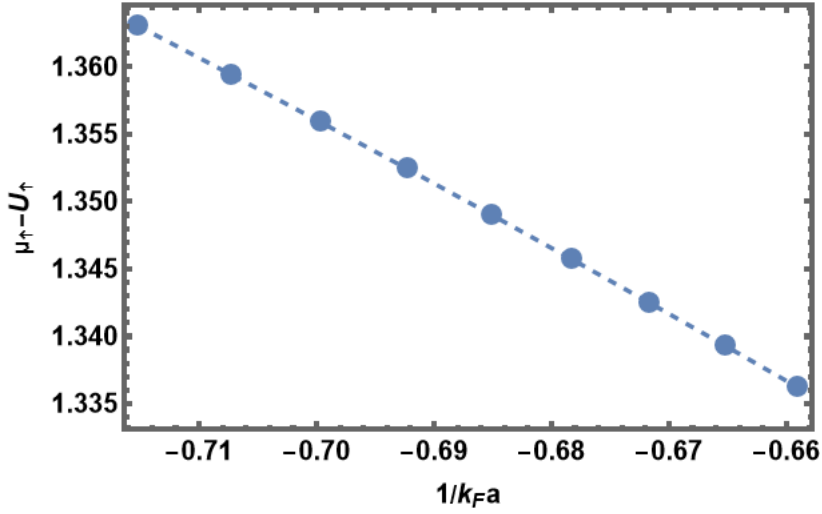

Fig. 9: (Color online) $\mu_\uparrow - U_\uparrow$ on the BCS side within the normal–superfluid phase separation regime, as a function of the interaction strength at a fixed mass ratio, $m_r = 3.7$.

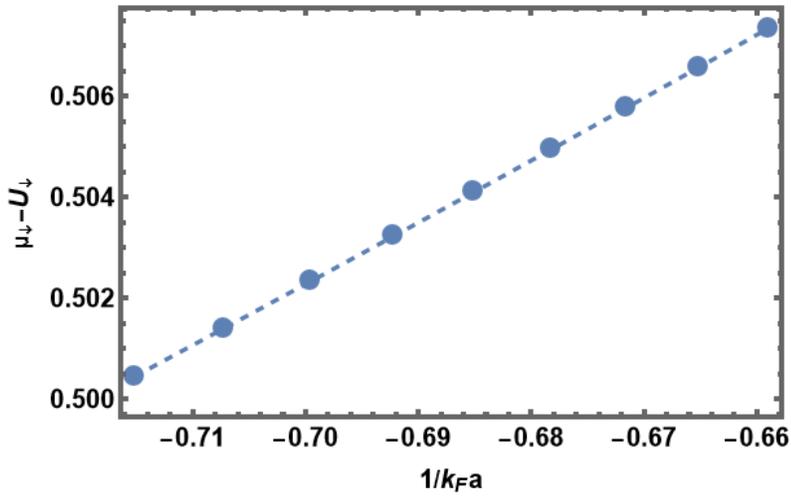

Fig. 10: (Color online) $\mu_\downarrow - U_\downarrow$ on the BCS side within the normal–superfluid phase separation regime, as a function of the interaction strength at a fixed mass ratio, $m_r = 3.7$.

Utilizing the obtained values for $h_s, \mu_s,$ and $\Delta$, the absorbed power can be determined.

## 3. Absorbed power
### 3-I. Method of Calculations

Now the time-dependent external perturbation is added to the system. For this purpose, and for wide class of perturbations, one can write the following perturbed Hamiltonian in momentum space[73-74]

$$H_1 = \sum_{\vec{k},\vec{k}',\sigma,\sigma'} B(\vec{k}\sigma|\vec{k}'\sigma') a^\dagger_{\vec{k},\sigma} a_{\vec{k}',\sigma'} \tag{22}$$

where $B(\vec{k}\sigma|\vec{k}'\sigma')$ is a general spin- and momentum-dependent potential matrix element. Some examples of $B(\vec{k}\sigma|\vec{k}'\sigma')$ are expressed as follows: (a) due to artificial vector potential $\vec{A}_{\vec{k}-\vec{k}'}$ and $B(\vec{k}\sigma|\vec{k}'\sigma')$ is proportional to $\vec{A}_{\vec{k}-\vec{k}'}\cdot(\vec{k}-\vec{k}')$, (b) due to sound waves and ultrasonic waves and $B(\vec{k}\sigma|\vec{k}'\sigma')$ can be proportional to deformation potential. Although, the external perturbation maybe cause to modulate the parameters such as energy gap, however, in this paper, it is supposed that the external potential is too weak to change the parameters. Also, the averaging on spin indices and wave vectors of $B(\vec{k}\sigma|\vec{k}'\sigma')$ can be taken in to account as $B(\vec{k}\sigma|\vec{k}'\sigma') = B = const$. Also, because of calculating the ratio of the absorbed power to that of the normal one, which is absorbed power with $\Delta = 0$ and $\mu_\uparrow = \mu_\downarrow$, $B$ is eliminated.

Now the perturbation Hamiltonian should be written in terms of Bogoliubov quasiparticle operators. The Bogoliubov transformation is defined by

$$a^\dagger_{\vec{k}\alpha} = u_{\vec{k}} \gamma^\dagger_{\vec{k}\alpha} + \sum_\beta \rho_{\alpha\beta} v_{\vec{k}} \gamma_{-\vec{k}\beta} \tag{23}$$

where $u_{\vec{k}}$ and $v_{\vec{k}}$ are the Bogoliubov coefficients and $\rho_{\alpha\beta}$ elements is given by

$$\rho = \begin{bmatrix} 0 & -1 \\ 1 & 0 \end{bmatrix} \tag{24}$$

$H_1$ can also been written in terms of $\gamma^\dagger$ and $\gamma$ as follows

$$H_1 = \sum_{\vec{k},\vec{k}',\sigma,\sigma'} B \left( u_{\vec{k}} u_{\vec{k}'} \gamma^\dagger_{\vec{k}\sigma} \gamma_{\vec{k}'\sigma'} + v_{\vec{k}} v_{\vec{k}'} \sum_{\beta,\beta'} \rho_{\sigma\beta} \rho_{\sigma'\beta'} \gamma_{-\vec{k}\beta} \gamma^\dagger_{-\vec{k}'\beta'} + \right.$$

$$\left. u_{\vec{k}} v_{\vec{k}'} \sum_{\beta'} \rho_{\sigma'\beta'} \gamma^\dagger_{\vec{k}\sigma} \gamma^\dagger_{-\vec{k}'\beta'} + v_{\vec{k}} u_{\vec{k}'} \sum_{\beta} \rho_{\sigma\beta} \gamma_{-\vec{k}\beta} \gamma_{\vec{k}'\sigma'} \right) \tag{25}$$

The term $\gamma^\dagger_{\vec{k}\sigma} \gamma_{\vec{k}'\sigma'}$ describes the scattering process of Bogoliubov quasiparticle from $|\vec{k}'\sigma'\rangle$ to $|\vec{k}\sigma\rangle$, and similarly for the term $\gamma_{-\vec{k}\beta} \gamma^\dagger_{-\vec{k}'\beta'}$. The term $\gamma^\dagger_{\vec{k}\sigma} \gamma^\dagger_{-\vec{k}'\beta'}$ ($\gamma_{-\vec{k}\beta} \gamma_{\vec{k}'\sigma'}$) also creates (destroys) two quasiparticles. One can then calculate the transition rate using Fermi golden rule at finite temperatures according to [73-74]

$$R_{\vec{k} \to \vec{k}'} = 2\pi \left| \langle \vec{k}'\sigma' | H_1 | \vec{k}\sigma \rangle \right|^2 \left\{ f(E_{\vec{k}_\alpha})(1 - f(E_{\vec{k}'_\alpha})) - f(E_{\vec{k}'_\alpha})(1 - f(E_{\vec{k}_\alpha})) \right\} \delta(E_{\vec{k}'_\alpha} - E_{\vec{k}_\alpha} - \omega) \tag{26}$$

where $f(E_{\vec{k}_\alpha})$ is the Fermi-Dirac distribution function, and $|\vec{k}\sigma\rangle$ is defined as

$$|\vec{k}\sigma\rangle \equiv a^\dagger_{\vec{k}\sigma} \prod_{\vec{k}' \neq \vec{k}} (u_{\vec{k}} + v_{\vec{k}} a^\dagger_{\vec{k}'\uparrow} a^\dagger_{-\vec{k}'\downarrow}) |0\rangle \tag{27}$$

or

$$\gamma^\dagger_{\vec{k}\sigma} |\psi_{BCS}\rangle = |\vec{k}\sigma\rangle \tag{28}$$

with $|\psi_{BCS}\rangle$ the generalized BCS state. By using Eqs. (25) and (28), the matrix element $\langle \vec{k}'\sigma' | H_1 | \vec{k}\sigma \rangle$ appeared in Eq. (26) becomes

$$\langle \vec{k}'\sigma' | H_1 | \vec{k}\sigma \rangle = B(u_{\vec{k}'} u_{\vec{k}} - \eta v_{\vec{k}'} v_{\vec{k}}) \tag{29}$$

where $\eta$ can be $\pm 1$ ($\eta = 1$ is used for attention low frequency sound wave). It should be mentioned that $\eta = 1$ for process which is even under time reversal. The absorption rate is also given by

$$\frac{dE}{dt} = W = \frac{1}{(2\pi)^6} \sum_{\alpha=1}^{2} \sum_{\sigma'=1}^{2} \iint d^3k \, d^3k' R_{\vec{k} \to \vec{k}'} \omega \tag{30}$$

Using Eqs. (26) and (29), and by changing from k- to E-space via $\sum_{\vec{k}} \ldots = (1/2\pi^2) \int k^2 dk \ldots = \int dE N_s(E) \ldots$, from Eq. (30), one can then write

$$W = 2\pi\omega|\overline{B}|^2 \int dEdE' N_s(E') N_s(E) \big((u(E')\ u(E) - \eta v(E')\ v(E))\big)^2 (f(E) - f(E'))\delta(E' - E - \omega)$$
(31)

where $N_s(E)$ is the density of states of the system and it is calculated using $N(E) = \sum_n \delta(E - E_n)$, with the delta function $\delta(E - E_n)$; the result is

$$N_s(E) = \frac{4\pi k(E)}{\dfrac{1}{m_-} + \dfrac{\left(\dfrac{1}{m_+}\right)\left(\dfrac{k^2}{2m_+} - \mu_s\right)}{\sqrt{\left(\dfrac{k^2}{2m_+} - \mu_s\right)^2 + \Delta^2}}}$$
(32)

where

$$k^2(E) = \frac{1}{\left(\dfrac{1}{(2m_-)^2} - \dfrac{1}{(2m_+)^2}\right)} \left(-\left[\dfrac{\mu_s}{2m_+} - \dfrac{E_\alpha + h_s}{2m_-}\right]\right.$$

$$\left. + \sqrt{\left[\dfrac{\mu_s}{2m_+} - \dfrac{E_\alpha + h_s}{2m_-}\right]^2 - \left(\dfrac{1}{(2m_-)^2} - \dfrac{1}{(2m_+)^2}\right)\left((E_\alpha + h_s)^2 - \Delta^2 - \mu_s^2\right)}\right)$$
(33)

### 3-II. Results and discussion

Fig. 11 illustrates the absorbed power(in arbitrary unit ) to that of the normal one, $W/W_N$, as a function of temperature measured by Fermi temperature, at a fixed interaction strength, the fixed external frequency(measured with respect to $T_F$) and two different mass ratios. $W_N$ refers to the normal noninteracting Fermi gas with $\Delta = h_s = 0$ and $m_r = 1$ and is given by $W_N = 2\pi\omega^2 N(0)^2 |B|^2$ where $N(0)$ is the density of states at Fermi surface [73-74]. As is seen, at a fixed temperature, the absorption rate decreases at lower mass ratios. The effect of mass ratio also becomes more important as temperature increases. The imbalance chemical potential, $h_s$, takes larger values with increase in mass ratio. By using $\mu_\uparrow = \mu_s + U_s + h_s$ and $\mu_\downarrow = \mu_s + U_s - h_s$ (all the relevant parameters are tuned here to make $\mu_\uparrow$ and $\mu_\downarrow$ positive), and by considering $U_s$ and $\mu_s$ are fixed, the value of $\mu_\uparrow$ is always greater than

$\mu_\downarrow$. As a result, the spin-down quasihole compared to the spin-up quasiparticle is then excited more easily, playing a more important role in the temperature dependence of absorbed power. Increase in mass ratio also increases the absorption rate. Of course, results may be altered when circumstances change. In Fig. 12, the temperature dependence of absorbed power at two different interaction strengths, a fixed mass ratio, and a fixed external frequency, is shown. Change in imbalance chemical potential, $h_s$, can be resulted from change in mass ratios or interaction strengths, and vice versa. Here, since the mass ratio is kept fixed, the two different interaction strengths then lead to different imbalance chemical potentials. In fact, keeping other parameters fixed and with $m_r = 3.7$, the interaction strength is then $1/k_F a \approx -0.659(-0.723)$, along with $h_s \approx 0.247\,(0.298)$. At a fixed temperature, the larger magnitude of interaction strength, the larger absorbed power. This is based on the fact that increase in $h_s$, or equivalently decrease in the absolute of scattering length, $|a|$, reduces the scattering. Therefore, the ultra sound wave as a perturbation can be absorbed better. From Figs. 11 and 12, at a high fixed temperature, and at a higher value of mass ratio (or at a higher value of the interaction strength), absorbed power becomes more noticeable. Fig. 13 illustrates the diagram of $W/W_N$ with respect to $h_s$ at two different values of mass ratio $m_r = 2.9$ and 3.0, a fixed temperature, $T/T_F = 0.0195$, and a fixed external frequency, $\omega = 0.1$, showing that $W/W_N$ increases with $h_s$, getting more noticeable at higher $h_s$. The effect of mass ratio is also more considerable at a high value of $h_s$.

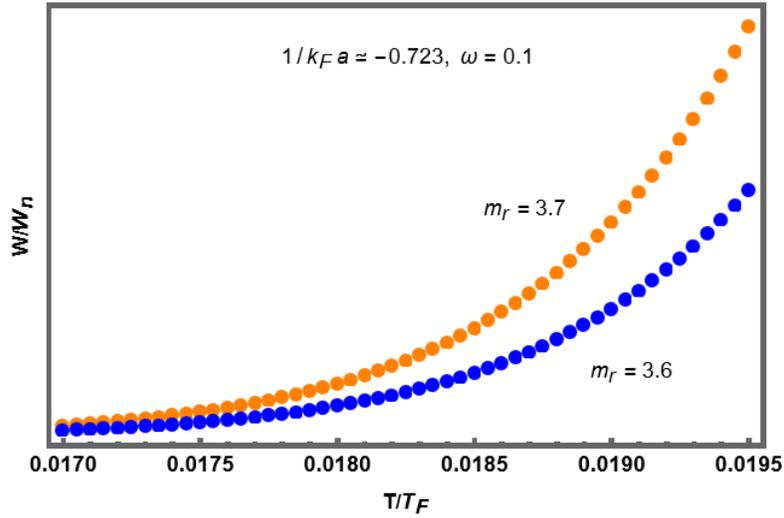

Fig.11: (Color online) Absorbed power (in arbitrary units) versus temperature, $T/T_F$, at a fixed interaction strength $1/k_F a = -0.723$ and a fixed external frequency $\omega = 0.1$ (measured with respect to $T_F$), and two different mass ratios $m_r = 3.6$ and $0.37$.

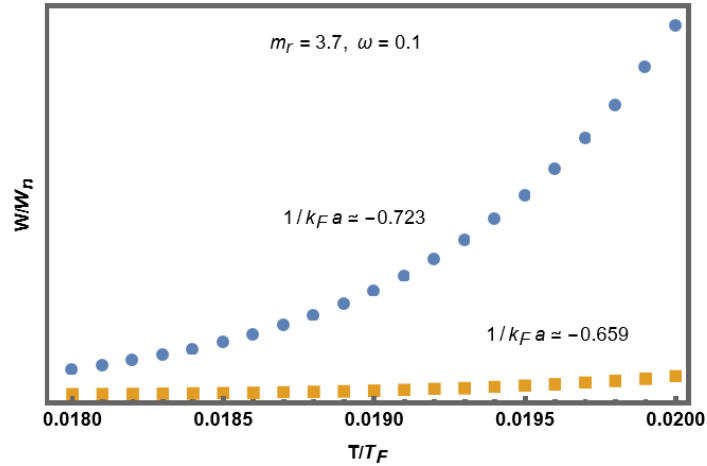

Fig. 12: (Color online) Absorbed power (in arbitrary units) versus temperature, $T/T_F$ at a fixed mass ratio $m_r = 3.7$ and a fixed external frequency $\omega = 0.1$ (measured with respect to $T_F$), and at two different interaction strengths $1/k_F a = -0.659$ and $-0.723$.

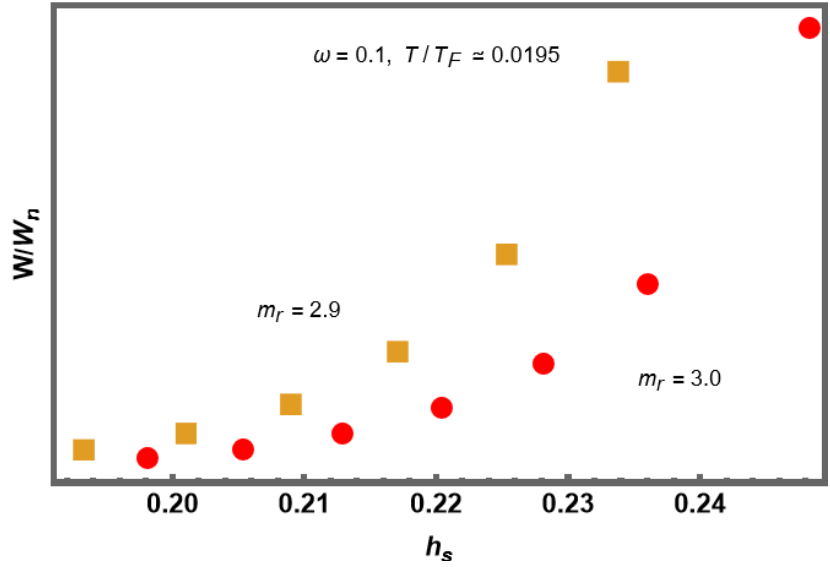

Fig. 13: (Color online) Absorbed power (in arbitrary units) versus imbalance chemical potential, $h_s$, at a fixed temperature $T/T_F = 0.0195$ and a fixed external frequency $\omega = 0.1$ (measured with respect to $T_F$), and at two different mass ratios, $m_r = 2.9$ and $m_r = 3.0$.

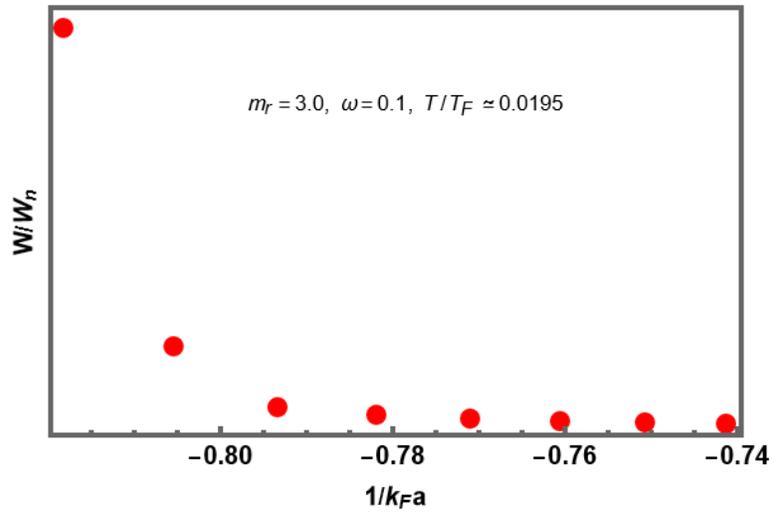

Fig. 14: (Color online) Absorbed power (in arbitrary units) versus interaction strength $1/k_F a$ at a fixed mass ratio $m_r = 3.0$, a fixed temperature, $T/T_F = 0.0195$ and a fixed external frequency $\omega = 0.1$ (measured with respect to $T_F$).

From Fig. 14, at a fixed mass ratio $m_r = 3.0$, a fixed temperature $T/T_F = 0.0195$, and a fixed external frequency $\omega = 0.1$, the absorbed power increases, when the magnitude of interaction strength increases.

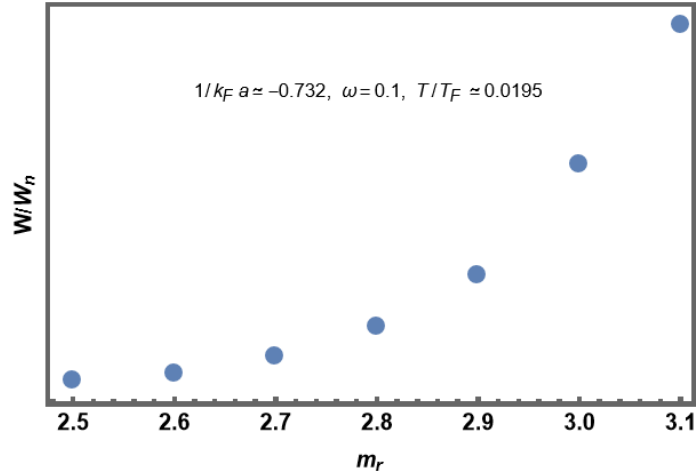

Fig. 15: (Color online) Absorbed power (in arbitrary units) versus mass ratio, $m_r$, at a fixed interaction strength $1/k_F a = -0.732$, a fixed temperature, $T/T_F = 0.0195$ and a fixed external frequency $\omega = 0.1$ (measured with respect to $T_F$).

From Fig. 15, it is inferred that at a fixed temperature $T/T_F = 0.0195$, a fixed interaction strength $1/k_F a = -0.732$, and a fixed external frequency $\omega = 0.1$, the effect of mass ratio on increase in absorbed power becomes more noticeable when mass ratio increases.

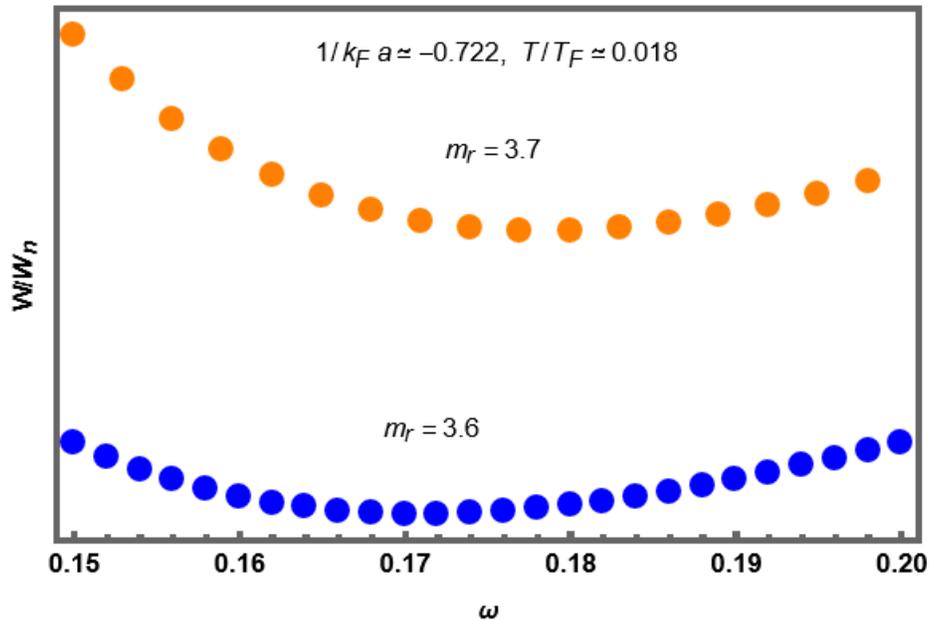

Fig. 16: (Color online) Absorbed power (in arbitrary units) versus external frequency $\omega$, (measured with respect to $T_F$) at a fixed $1/k_F a = -0.722$, a fixed temperature $T/T_F = 0.018$ and two different mass ratios, $m_r = 3.6$ and $3.7$.

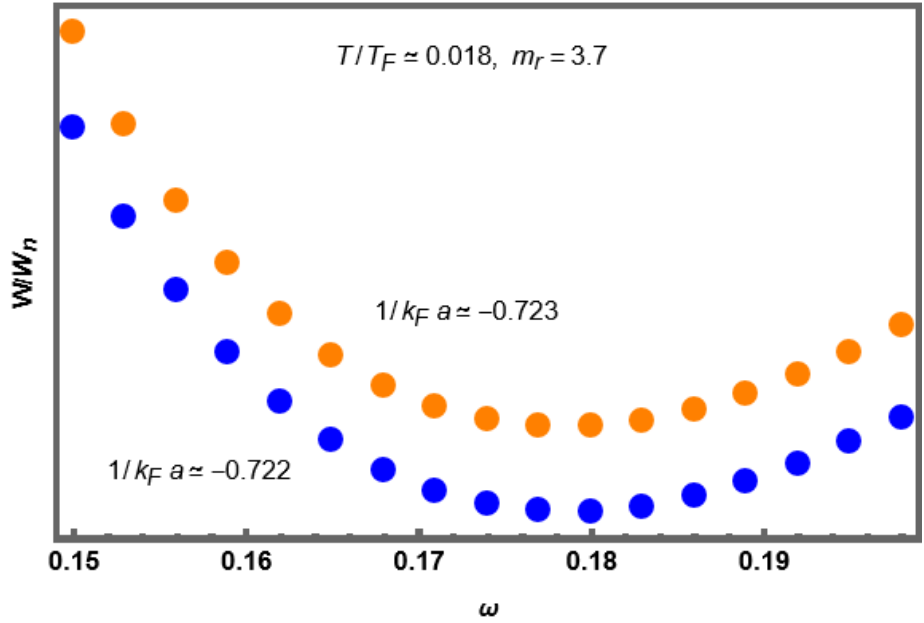

Fig. 17: (Color online) Absorbed power (in arbitrary units) versus external frequency $\omega$ (measured with respect to $T_F$) at fixed mass ratio $m_r = 3.7$, a fixed temperature $T/T_F = 0.018$ and two different interaction strengths, $1/k_F a = -0.722$ and $-0.723$.

From Fig. 16 (Fig. 17), it is seen that at a fixed frequency, for Fermi-Fermi mixtures with a higher mass ratio (the magnitude of interaction strength), the absorbed power is higher. It should be mentioned that at a fixed mass ratio (the magnitude of interaction strength), the absorption at a special frequency has a minimum due to the structure of density of states of the system.

## 4. Conclusions

The absorbed power perturbed by ultrasound wave subject to a polarized Fermi-Fermi mixture composed of spin-down quasiholes and spin-up quasiparticles was calculated in BCS side of BCS-BEC crossover and in the presence of normal-superfluid phase separation . Different masses as different Fermi-Fermi mixtures were examined. However, prior to the investigation of the absorbed power, it was necessary to provide information on the variations of the key quantities. Dependencies of the average and imbalance chemical po-tentials on the interaction strength and on the polarization of the normal component, together with the order parameter or energy gap function versus interaction strength, were determined. In particular, the imbalance chemical potential and the polarization were obtained as functions of the mass ratio. These results identified the ranges of parameters that preserved normal–superfluid phase separation and enabled an estimate of the transition temperature to the fully normal state, which required that the parameters used in the absorbed power calculation be below this temperature. The results showed that, with the other parameters held fixed, both a larger interaction strength (or equally the average chemical potential) and a larger mass ratio increased the imbalance chemical potential.

On a microscopic level, mass imbalance strengthens the suppression of pairing formation and consequently affects the occurrence of normal–superfluid phase separation, by increasing the mismatch between the two spin components' Fermi surfaces and by altering the density of states for each component. The results show that, as the mass ratio increases, a larger imbalance chemical potential is required for the thermodynamic conditions of normal–superfluid phase separation to be satisfied. Increasing the imbalance chemical potential is an additional energy cost that stabilizes the system by placing the unpaired population in the region surrounding the superfluid component, which minimizes the energy of the superfluid component. At the same time, the required increase in the imbalance chemical potential is accompanied by a decrease in both the total polarization and the polarization of the normal component; equivalently, the spin-population imbalance in the normal component is reduced.

At fixed temperature, interaction strength, and external frequency, Fermi-Fermi mixtures with higher allowed mass ratio exhibit a more significant change in the absorbed power. Additionally, as the mass ratio increases, the absorbed power also increases. Since the frequency of the ultrasonic wave is lower than the pair-breaking threshold, the absorption is mainly governed by the quasiparticles and quasiholes present in the system. When the mass ratio increases, the mass of the spin-down quasiholes becomes larger, making them heavier and thus reducing their average velocity. As a result, their collision rate with the wave increases, enhancing the probability of energy absorption, which contributes to the overall increase in absorption power. Moreover, an increase in the mass ratio alters the energy spectrum of the spin-down quasiholes, making it more difficult for them to form Cooper pairs with spin-up quasiparticles. This reduction in pairing tendency leads to a higher population of unpaired population in the normal component, which further facilitates energy absorption. Altogether, these effects result in the observed increase in absorption power with increasing mass ratio. Moreover, the effect of mass ratio becomes more considerable as temperature increases. For a Fermi-Fermi mixture, an increase in the magnitude of interaction strength or the imbalance chemical potential leads to an enhancement in the absorbed power. At a fixed mass ratio or a fixed magnitude of interaction strength, There is a frequency at which the absorption power is minimized.

## Data Availability Statement

All the data on which the present investigation relies are present in the manuscript.

## Aknowledgement

I thank the anonymous reviewer for their valuable comments, which have greatly contributed to improving this manuscript.